\documentclass[a4paper]{jpconf}
\usepackage{graphicx}
\usepackage{url}

\def\arcmin{\hbox{$^\prime$}}

\def\HII{H\,\textsc{ii}}
\def\ha{H$\alpha$}

\def\NII{[N\,\textsc{ii}]}

\begin{document}
\title{Discovery of new planetary nebulae in the Small Magellanic Cloud}

\author{D. Dra\v{s}kovi\'{c}$^1$, Q. A. Parker$^{2,4}$, W. A. Reid$^{1,3}$ and M. Stupar$^3$}

\address{$^1$ Department of Physics and Astronomy, Macquarie University, NSW, Australia\\
$^2$ Department of Physics, The University of Hong Kong, Pokfulam Road, Hong Kong  \\
$^3$ School of Computing, Engineering and Mathematics, Western Sydney University, Australia \\
$^4$ Australian Astronomical Observatory (AAO), North Ryde, NSW, Australia}

\ead{$^1$ danica.draskovic@mq.edu.au}

\begin{abstract}
We present six new planetary nebulae (PNe) discovered in the Small Magellanic Cloud (SMC) from deep UK Schmidt telescope (UKST) narrow band \ha{} and broad-band short-red ``SR" continuum images and confirmed spectroscopically. These 6 preliminary discoveries provide a 6\% increase to the previously known SMC PN population of $\sim$100. Once spectroscopic follow-up of all our newly identified candidates is complete, we expect to increase the total number of known SMC PNe by up to 50\%. This will permit a significant improvement to determination of the SMC PN luminosity function (PNLF) and enable further insights into the chemical evolution and kinematics of the SMC PN population.
\end{abstract}

\section{Introduction}

Planetary nebulae (PNe) play a crucial role in understanding mass loss for low and intermediate mass stars, they influence chemical evolution of galaxies through interstellar enrichment and are a major Galactic dust factory. Their strong emission lines are laboratories for plasma physics and provide accurate radial velocities for Galactic kinematics  while
the PNLF  in external galaxies is a powerful cosmological distance indicator.

 \noindent The SMC  is a gas rich, late type dwarf irregular galaxy with an average gas/dust ratio a factor of $\sim30$ below the Galactic value \cite{SS2000}. The SMC's accurately determined distance of 61$\pm1$~kpc \cite{HHH2005} allows derivation of true PNe physical parameters from optical spectra. With a  small angular extent of $5^{\circ}20\arcmin\times3^{\circ}05\arcmin$, it can be easily studied in its entirety. SMC reddening and extinction is relatively low and uniform, enabling absolute nebula luminosity estimates.
 
  \section{Discovery technique}
We have carefully examined UKST \ha\ and SR SMC image data for $\sim$120 $\deg^2$. We adopted the same powerful and proven colour merging technique successfully applied to the LMC by \cite{RP2006a}\cite{RP2006b}\cite{RP2013}. Using the \textsc{KARMA} visualisation software package, we surveyed these fields and created merged false-colour images, combining \ha\ (coloured in blue) and SR (coloured in red). All images were systematically visual scanned, looking for a faint, compact or barely resolved objects. By carefully choosing the image combination parameters we could balance the intensity of \ha\  and SR red matching images, allowing only specific features of one or other pass-band to be seen. In our scheme emission objects such as PNe and \HII\ regions appear with a uniform strong blue colour, whereas normal continuum stars are a uniform pink-purple colour. Emission-line stars usually have a strong continuum component and so are easily detected by the narrow extent of their blue halos around a strong pink core \cite{RP2006a}\cite{RP2006b}.

\begin{figure}
\begin{center}
\includegraphics[width=100mm]{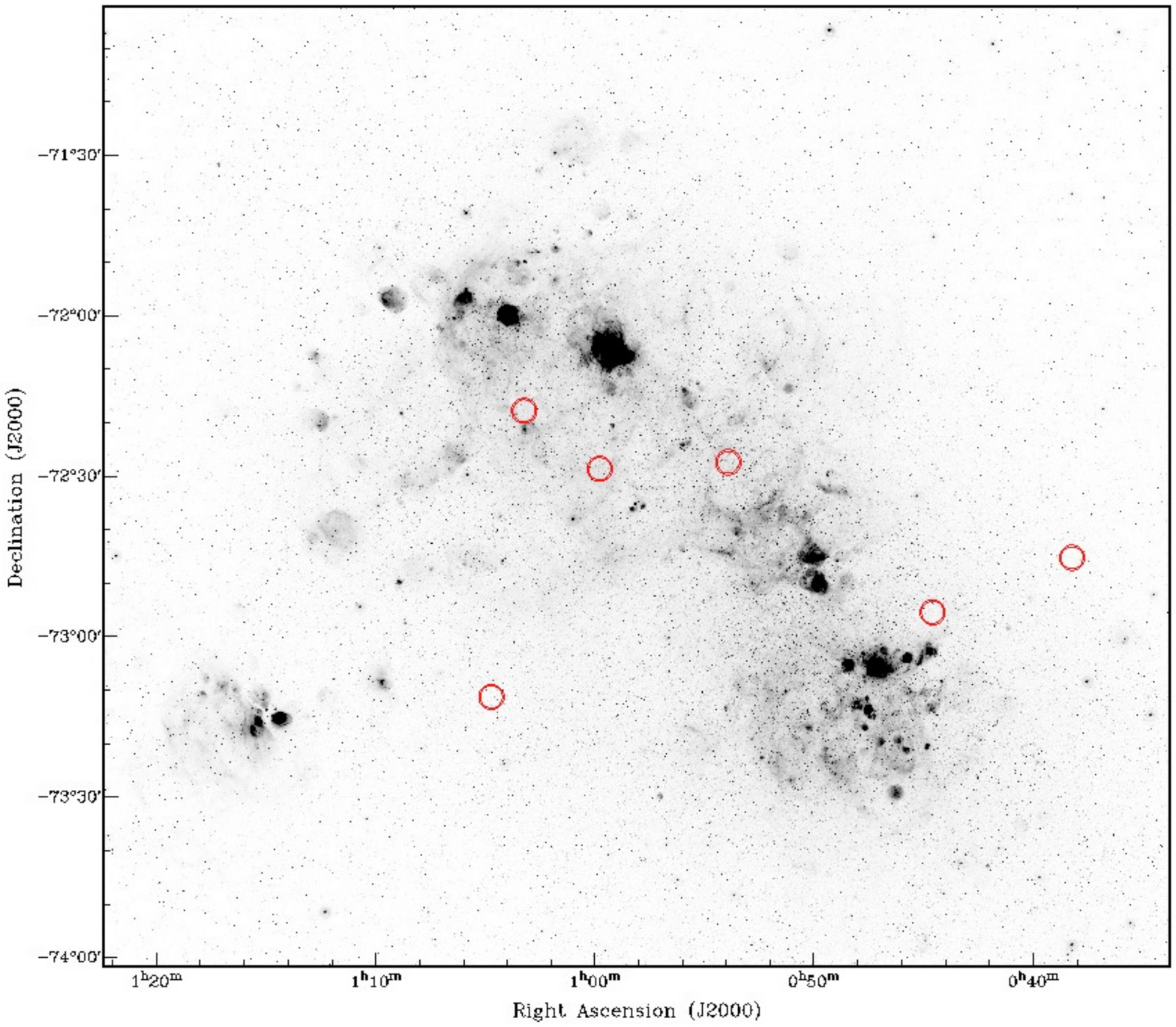}
\end{center}
\caption{\footnotesize{Positions of six new PNe discovered in the SMC.}}
\end{figure}

\section{Conclusions}
Confirmatory spectra of six preliminary PNe candidates are quite homogenous. Each was confirmed as an emission-line source with no obvious stellar continuum. This eliminates confusion with emission-line stars for compact sources. Each candidate also gave high \NII\ to \ha\ ratios ($\sim$ 2 to 7) \cite{DPRS2015} not observed in \HII\ regions, which are the most likely contaminants \cite{KBFM2000}. The presence of only narrow diagnostic lines also removes the confusion with supernova remnants. These preliminary results demonstrate the strength and promise of our techniques for providing high-quality PNe candidates. 

We expect to increase SMC PNe by  20 to 50\% after spectroscopic follow-up of our candidates and refinement of identification of all known SMC emission-line sources. 
This will include multi-wavelength analysis and spectroscopic confirmation (or rejection) of objects previously listed as SMC PNe. Our survey will permit more extreme ends of the PNLF to be explored, especially at the faint end, and enable us to study underrepresented evolutionary stages of SMC PNe. This sample will have a major impact on improving rates of elemental enhancement in lower mass stars within a low-metallicity environment. Other advances will come from the improved SMC kinematical data from over 300 emission-line objects uncovered.

\end{document}